\documentclass[oribibl]{llncs}

\usepackage{paralist}
\usepackage{rotating}
\usepackage{wrapfig}
\usepackage{amsfonts}	
\usepackage{url}
\usepackage{hyphenat}
\usepackage{nth}
\usepackage{xspace}
\usepackage{mathpartir}

\usepackage{algorithmicx}
\usepackage[noend]{algpseudocode}
\usepackage{color}
\usepackage{amsmath}
\usepackage[bookmarks]{hyperref}
\usepackage[capitalize]{cleveref}

\usepackage{tikz} 
\usetikzlibrary{shapes.geometric, arrows}

\DeclareMathOperator*{\argmax}{\arg\!\max}
\DeclareMathOperator*{\argmin}{\arg\!\min}

\newcommand{\pdftitle}{Information-theoretic User Interaction: Significant Inputs for Program Synthesis}                     
\newcommand{\pdfkeywords}{entropy, program synthesis, synthesis by examples, active learning, distinguishing input, clustering}
\newcommand{\pdfsubject}{Program Synthesis}

\hypersetup{
	pdftitle={\pdftitle},
	pdfauthor={SG, DP, AR, AT},
	pdfkeywords={\pdfkeywords},
	pdfsubject={\pdfsubject},
	pdflang={en-US}
}
\title{\pdftitle\thanks{%
A note on history: This article was submitted to CAV 2018, 2019, and 2020, and PLDI 2019 and 2020. 
A recent paper in PLDI 2020 titled ``Question selection for interactive program synthesis'' 
(independently) addresses the same problem.
}}
\titlerunning{Information-theoretic User Interaction}
\author{Ashish Tiwari, Arjun Radhakrishna, Sumit Gulwani, and Daniel Perelman}
\institute{
  Microsoft, Redmond, WA
}

\begin{document}
\newcommand{\ignore}[1]{}
\newcommand{\arsays}[1]{\textbf{AR: } #1}
\newcommand{\ap}[1]{{\color{red}\textbf{AP:} #1}}

\def\la{{\langle}}
\def\ra{{\rangle}}
\def\ZZ{{\mathbf{Z}}}

\def\SamplingSpec{\mathtt{SamplingSpec}}

\newcommand{\spec}{\ensuremath{\varphi}\xspace}

\newcommand{\numberOfBenchmarks}{791\xspace}

\maketitle

\begin{abstract}
Programming-by-example technologies are being deployed in
industrial products for real-time synthesis of various kinds of data transformations.
These technologies
rely on the user to provide few representative examples of the transformation task.
Motivated by the need to find the most pertinent question to ask the user,
in this paper, we introduce the {\em significant
  questions problem}, and show that it is hard in general.
  We then develop an information-theoretic greedy approach for 
  solving the problem. We justify the greedy algorithm using the conditional
entropy result, which informally says 
  that the question that 
  achieves the maximum information gain is the one that we know least about.

In the context of interactive program synthesis, we use the above result to
    develop an {\em{active program learner}} that
    generates the significant inputs to pose as queries to the user in each iteration.  
The procedure requires extending a {\em{passive program learner}} to a {\em{sampling program learner}} that is able
to sample candidate programs from the set of all consistent programs to enable estimation of 
information gain. 
It also uses clustering of inputs based on features in the inputs and the corresponding outputs
to sample a small set of candidate significant inputs.
Our active learner is able to tradeoff false negatives for
false positives and converge in a small number of iterations on a real-world dataset of 
string transformation tasks. 
\end{abstract}

\section{Introduction}
In recent years, the field of \emph{automatic program synthesis} of \emph{data transformation programs} from
user-provided \emph{example-based specifications} has received wide attention from the industrial community~\cite{pbe-wrangling}.
Data transformation programs commonly arise in machine learning~\cite{wrangler}, healthcare~\cite{noodles}, and IT
administration~\cite{flashextract}, as well as in any business analytics that involves \emph{data wrangling}.
The ability to process data by providing examples of the desired transformation not only makes data wrangling 
approachable to non-programmers, but also simplifies data scientists' workflows, who commonly
dedicate as much as 80\% of their time to manual wrangling~\cite{wrangler}.
This has led multiple software companies to incorporate semi-automatic transformation synthesis in their machine
learning IDEs, including Microsoft's Azure ML
Workbench~\footnote{\url{https://azure.microsoft.com/en-us/services/machine-learning-services}}, Google's Cloud
Dataprep~\footnote{\url{https://cloud.google.com/dataprep}} (based on Trifacta\footnote{\url{https://trifacta.com}}),
Uber's Michelangelo~\footnote{\url{https://eng.uber.com/michelangelo}}, and
Tableau~\footnote{\url{https://www.tableau.com/}}.

Specifying programs by input-output examples is notoriously ambiguous.
Even an extensively engineered program synthesis system may require
as many as 7 examples to correctly identify an intended program~\cite{flashprog}.
In the past, intent ambiguity has typically been addressed by trying to heuristically avoid it: 
impose a sophisticated ranking on the underlying
domain-specific language~\cite{cav:ranking} so that ranking disambiguates user intent. 
However, heuristics can fail, and in such cases,
the user is responsible for finding an input where the synthesized transformation does not match
their intent and provide an additional example.
This is relatively straightforward when datasets are small where the user can simply eye-ball the
data~\cite{flashfill}, but much more challenging with a large dataset intended for business analytics.

The intent ambiguity challenge has caused the industry to embrace an \emph{interactive} and \emph{predictive} approach
to program synthesis.
Synthesis proceeds in rounds wherein the system proactively makes suggestions to the user, and the user provides information accordingly.
For instance, Azure ML Workbench suggests a subset of \emph{significant inputs} from the data that may best disambiguate
the hypothesized transformations~\cite{derivecolumn},
and Trifacta Wrangler suggests possible next steps in the desired transformation~\cite{trifactapredictive}.

While the academic community has long modeled program synthesis as an iterative interactive process~\cite{ogis,peleg18},
proactively generating high-quality examples, or constraints, to optimize convergence to the intended program is
still an open problem.
With the concrete goal of building an {\em{active program synthesizer}},
we start by formulating the problem of generating optimum user queries in an abstract setting. 
We show its hardness, and
then cast the problem in a probabilistic framework to enable application of information-theoretic methods.
We then use the chain rule for conditional entropy to design a greedy algorithm for the problem.
Intuitively, the chain rule says that the system can pick the question whose answer will 
yield the most information gain by finding the question about which the system knows the least.

The abstract greedy information-gain procedure is instantiated to build an active program learner. 
The active program learner iteratively refines its belief of the intended program.
This belief state is just a probability distribution over the program space.
We now face two challenges. First, working with
probability distributions over program space is intractable.
We overcome this challenge by estimating probability distributions using
Monte-Carlo methods.
In fact, we describe a {\em{sampling program learner}} that uses importance sampling to generate a
belief state (and not just one correct program).
Second, the number of inputs (that is, the number of possible questions we can ask the user)
can be really large.  We address this challenge by presenting a clustering-based approach for sampling of 
input space, where our key idea is to use features from both
the input, and the output corresponding to that input.

We evaluate our active program learner based on whether it
(a) minimizes the number of synthesis iterations, (b) minimizes the
number of \emph{false positive queries} (\emph{i.e.} extraneous queries when the learned program is already
correct), and (c) minimizes the number of \emph{false negative queries} (\emph{i.e.} missing queries when the
learned program is actually incorrect).
These criteria are difficult to satisfy simultaneously, but we present a comprehensive evaluation of different
techniques to pick a trade-off solution.

\section{Significant Questions}\label{sec:significantquestions}


Consider a blackbox software system $bb$. 
Let us say we want to answer a fixed question $q$ about $bb$.
To answer the question $q$, we can ask questions from a predefined set 
$QS = \{q_1, q_2, \ldots, q_n\}$ of questions, and we assume there is an oracle
(say, a user) that can answer these questions about $bb$.
We are interested in the following problem: which of the $n$ questions
should we ask the oracle?
Our goal is to minimize the number of interactions with the oracle required 
in the process of answering the question $q$ about the given system $bb$. 

The hypothesis space, $HS$, is a set $\{p_1, \ldots, p_N\}$ of all possible values that
$bb$ can take; in other words, $bb$ is known to belong to the set $HS$.
In general, the set $HS$ need not contain the concrete programs, but only some abstractions
that are sufficient to answer the questions $q$ and $q_1, \ldots, q_n$. 
This distinction is not important here, so for simplicity,
assume that $HS$ contains concrete programs.

The answer space, $AS$, is a set $\{a_1, \ldots, a_m\}$ of all possible answers for the questions in $QS$.
A given question-answer pair, $(q_i, a_j)$, can either be consistent with a given hypothesis $p_k$,
or inconsistent with it.
The notation $p_k \models (q_i, a_j)$ denotes that hypothesis $p_k$ is consistent with $(q_i, a_j)$,
and 
$p_k \not\models (q_i, a_j)$ denotes it is not.

We are interested in finding a {\em{plan}} for asking questions. A {\em{plan}} is a 
mapping $\sigma: (QS\times AS)^* \mapsto QS\cup\{\bot\}$ that maps a history of question-answer
pairs, possibly of length $0$, to the next question to ask. 
A plan $\sigma$ is {\em{terminating}} if there is a finite number
$k$ such that $\sigma( (QS\times AS)^{k'} ) = \bot$ for all $k' \geq k$.


A sequence of question-answer pairs, 
\begin{eqnarray}
    (q_0,a_0), (q_1,a_1), \ldots, (q_l,a_l), \label{eqn:ioseq}
\end{eqnarray}
is {\em{consistent}} (with respect to a plan $\sigma$ and a program $bb$) if
\\
(a) $bb \models (q_i,a_i)$ for all $i=0,\ldots,l$; that is, each answer $a_i$ correctly answers the
question $q_i$ about the program $bb$, and
\\
(b) $q_{i+1} = \sigma(\langle (q_0,a_0),\ldots,(q_i,a_i)\rangle)$; that is, the questions in the sequence
are picked using the given plan.

A feasible sequence of question-answers, as in~(\ref{eqn:ioseq}),
is 
{\em{maximal}} (with respect to terminating plan $\sigma$) if 
$\sigma(\langle(q_0,a_0),\ldots,(q_l,a_l)\rangle) = \bot$.

Maximal feasible sequences are important to state the correctness requirement of any plan:
Given any 
maximal feasible sequence of question-answer pairs, we should be able to deduce enough about the
unknown program $bb$ to 
answer the question
$q$ about it.

\begin{definition}[Significant Questions Problem]\label{def:prob}
    Given a hypothesis space $HS = \{p_1, \ldots, p_N\}$,
    a set $QS = \{q_1, \ldots, q_n\}$ of questions, 
    a set $AS = \{a_1, \ldots, a_m\}$ of answers, 
    synthesize a terminating plan $\sigma: (QS\times AS)^* \mapsto QS\cup\{\bot\}$ s.t. given any 
    sequence of the form
    $$
     (q_0,a_0), (q_1, a_1), \ldots, (q_l, a_l)
    $$
    that is maximal and feasible with respect to the plan $\sigma$ and program $bb$,
    it is possible to deduce an ``$a$'' s.t. $bb \models (q,a)$.
\end{definition}

Each plan $\sigma$ can be visualized as a tree:
each node in the tree is labeled with a question,
each node has as many children as there are answers to its question,
the root node is labeled with $\sigma(\epsilon)$,  and every other node is 
labeled by the question generated by the plan $\sigma$ based on the question-answer pairs 
on the path from the root to that node.
The process of 
answering $q$ about $bb$ using the plan $\sigma$
corresponds to traversing a path from a root to a leaf in the tree for $\sigma$.

The {\em{optimum significant questions}} problem seeks to find a plan that has a minimum
value for the worst-case number of questions asked.  In terms of the tree visualization,
we want the plan whose tree has the least height.

\begin{example}[Inserting element in a sorted list]\label{example1}
We can cast the problem of inserting a number into a given sorted list of $n$ numbers as a 
    optimum significant questions problem.
The unknown $bb$ here is the {\em{input}} number that has to be inserted (in a known sorted list).
The question $q$ we want answered about $bb$ is:
what is the {\em{position}} where we need to insert $bb$.
The possible answers $a$ for this question $q$ are $\{0, 1, \ldots, n\}$.
    The set of all questions $QS$ we are allowed to ask are
$\{q_0, \ldots, q_n\}$, where $q_i$ asks if
$a \leq i$, and the set of possible answers $AS$ is $\{true, false\}$.
The goal is to find the answer $a$ by asking the fewest number of questions
(in the worst case).
An optimum plan here would correspond to the binary search procedure: the first question would be
    $q_{\lfloor\frac{n}{2}\rfloor}$, and depending on the answer, the next question would be
    $q_{\lfloor\frac{n}{4}\rfloor}$ or
    $q_{\lfloor\frac{3n}{4}\rfloor}$, and so on.
The tree visualizing the binary search plan has height $O(\log(n))$.
\end{example}

\section{Information-Guided User Interaction}\label{sec:information}


In this section, we present  a greedy approach for solving the optimum signficant questions problem.
First, we note that achieving optimality is NP-hard: this follows by a reduction from set cover.
Hence, we resort to greedy methods. But, before we describe our greedy
approach, we need to cast the problem in a general probabilistic framework.

Let $E$ be a set of question-answer pairs.
The unknown artifact $bb$ can be viewed as a  random variable that
can take one of the values in $HS$.
Let $Pr(bb=p_k \mid E)$ denote the probability that the blackbox program $bb$ is $p_k$
given that $bb$ is known to be consistent with all the question-answer pairs in $E$.
Clearly, $Pr(bb \mid E)$ is a probability distribution over the hypothesis space $HS$.

We can answer any given question about $bb$ if
we know the identity of $bb$. 
For the rest of the paper, we shall assume that the question $q$ we want to answer about $bb$
is just the identity of $bb$, and we will use $q$ as a variable that ranges over the questions 
in $QS$ that we can ask the oracle.
Our knowledge about the identity of $bb$ is 
directly measured by the {\em{entropy}}
$En(Pr(bb)) = \sum_k -Pr(bb=p_k)\log(Pr(bb=p_k))$ of the probability distribution $Pr(bb)$.
Clearly, our goal is to reduce the entropy of $Pr(bb)$.

Initially, we do not have answers to any questions, and hence $E = \emptyset$, and our belief
about $bb$ is given by the probability distribution $Pr(bb \mid E=\emptyset) = Pr(bb)$.
Now, assume we ask question $q$ from $QS$.
Let $Pr(bb \mid q)$ denote the probability distribution on $bb$ conditioned on knowing the
answer to the question $q$. 
We view $q$ also as a random variable which takes values in the answer set $AS$.
Using the chain rule for conditional entropy, we can compute the entropy $En(Pr(bb\mid q))$ of
the distribution we get {\em{after asking the question $q$}} as
\begin{eqnarray}
    En(Pr(bb \mid q)) & = & En(Pr(bb)) - En(Pr(q)) \label{eqn:en}
\end{eqnarray}
where $Pr(q)$, a probability distribution on the answer space $AS$, is defined by
\begin{eqnarray}
    Pr(q = a) & = &  \sum_{\{p \mid p\models (q,a)\}} Pr(bb = p)
\end{eqnarray}

In the information-theoretic user interaction model, we solve the {\em{significant questions problem}}
by choosing the next question $q$ so that it (greedily) minimizes the entropy of $Pr(bb\mid q)$.
Equation~\ref{eqn:en} shows that the most greedy choice would be the question $q$ whose entropy
$En(Pr(q))$ is maximum.
We can generalize the observation of Equation~\ref{eqn:en} to the case when we have already
obtained answers to some $i$ prior questions.
If $E$ denotes a sequence $(q_1,a_1),\ldots,(q_i,a_i)$ of question-answer pairs that have been
obtained so far, then the 
{\bf{greedy plan}} $\sigma^*$ is given by:
\begin{eqnarray*}
    \sigma^*(E) & = & \left\{ \begin{array}{ll}
        \bot & \mbox{ if $En(Pr(q_E^* | E)) = 0$}
        \\
        q_E^* & \mbox{ otherwise}
    \end{array}\right.
\end{eqnarray*}
where $q_E^* = \argmax_{q}\{ En(Pr(q \mid E)) \}$.

The following proposition says that the greedy plan is indeed greedy.
\begin{proposition}\label{prop:main}
    Consider an instance of the significant questions problem where the question $q$ is
    the same as identity of $bb$. 
    If this instance has a solution, then the greedy plan $\sigma^*$ will be a solution.
    Although the plan $\sigma$ may not be optimum, it is greedy in every step; that is, 
    whenever $\sigma^*(E) \neq\bot$,
    it is the case that $\sigma^*(E) = \argmin_q\{ En(bb\mid E,q)\}$.
\end{proposition}

 Since the knowledge we have about a random variable is inversely related to its entropy,
 the above result intuitively states that, to greedily seek knowledge, we should ask
 the question about which we know the least.  
 
           \begin{example}\label{example2}
               Continuing with Example~\ref{example1},
               assume that the prior probability distribution (for the index $q$ where the unknown input
               will be inserted in the sorted list)
               is a uniform distribution over $\{0,\ldots,n\}$;
               that is,
               $Pr(q=i) = 1/(n+1)$ for all $i$.
               If we want to minimize the entropy,
               we know that the optimum choice would be question $q_i$
               that maximizes $En(q_i)$. 
               Now, the question $q_i$ has two possible answers, and hence the possible answers for $q$
               are partitioned into two clusters,
               namely, $\{0, 1, \ldots, i\}$ where the answer is true, and
               $\{i+1, \ldots, n\}$ where the answer is false. Hence, we have
               $$
               \sigma^*({\emptyset}) =  \argmax_{q_i\in QS}  -\frac{i+1}{n+1} \log(\frac{i+1}{n+1})  - \frac{n-i}{n+1}\log(\frac{n-i}{n+1})
               $$
               We know this entropy is maximized when $i$ is the floor of $(n+1)/2$.
               Thus, we get the binary search procedure.
               Note that the greedy approach 
               also suggests a way to generalize binary search when the
               (prior) distribution of elements to be inserted is not uniform.
           \end{example}

           \ignore{ 
\begin{remark}\label{remark:q}
    The definition of $En$ given above measures how much
    uncertainty there is in the exact identity of $bb$.  We can change the definition to measure how much 
	uncertainty we have
    about the answer to the specific question $q$ about $bb$ (and not necessarily know the value of $bb$).
    We will discuss this aspect later.
\end{remark}
 \endignore}

\section{Active Program Synthesis}
\label{sec:illustrative}

\ignore{
  Structure:
  \begin{itemize}
    \item The example setting
    \item First input-output example and top-K of programs
    \item Inputs on which multiple top programs differ
    \item Brief algorithm outline
    \item Justification for random programs
    \item Justification for input clustering and output clustering
  \end{itemize}

Scenario 1: There is a text extraction task already in sec 2.

Scenario 2: Concatenation of a bunch of tokens where each one can be extracted in two ways
“123 ABC 456 DEF … “  -> “123456…”
Now if there are two highly and closely ranked sub-programs for extracting each sequence of numbers, all the top programs will be combinations of those.
In this example, if the two closely ranked sub-programs are “take 3 digits” and “take all digits”, all top ranked program will be of the form
P1: Concat(Take 3 digits from beginning of string, Take 3 digits following second space, …)
P2: Concat(Take 3 digits from beginning of string, Take all digits following second space, …)
P3: Concat(Take all digits from beginning of string, Take 3 digits following second space, …)
P4: Concat(Take all digits from beginning of string, Take all digits following second space, …)

Scenario 3: The CPT example that appends a closing brace only if it is not present;
a.	Also, we can use an unit standardization task. Say “12 in” -> “12”, “8 in” -> “8”, “30 cm” -> “?”.
Even though the programs are unlikely to distinguish between “12 inch” and “30 cm”, the input formats are different and “30 cm” should be a significant input.
b.	We should make the point that if there are many different input patterns, there are likely to be too many significant inputs.

Scenario 4: Extracting the year from dates formatted in many different forms. (say 05-Dec-2015, 5 December 2015, 2015-12-5, 12/5/2015, etc)
                Input clustering will give too many significant inputs; however, output pattern should be very simple (4 digits)

Also, here are some general notes:
1.	Clustering good because it looks at input structure
2.	Clustering bad because it ignores task at hand:
a.	Extracting area code from differently formatted phone numbers gives many significant inputs if you consider input formats, but is an easy task (take first 3 digits)
3.	TopK + RandomK program based distinguishing inputs good as it looks at task at hand
4.	TopK + RandomK distinguishing bad as it ignores input structure

}

The main application of the greedy approach for user interaction design
we pursue in this paper is the
significant input problem in interactive program synthesis.
The key challenge in implementing the greedy information gain procedure
is to find ways to estimate the entropy of the different questions that
can be posed to the user.  Since the number of programs and the number
of inputs can be very large, we use sampling techniques to estimate the
various probabilities for computing entropies. 

\ignore{
We motivate the problem and our solution using the example data from
Figure~\ref{fig:illustrative_example}.
The figure shows semi-structured descriptions of an
inventory, where the user is attempting to extract the product identifiers from the
descriptions, as depicted in the ``Desired output'' column.
The user is unable to write the required program, but can provide the desired output on 
any given input.

The user starts the process of synthesizing the required data transform by
providing the output for the first input to the synthesizer, i.e., ``12
units PID 24122 Laptop'' $\mapsto$ ``PID 24122''.
Given this input-output example, typical example-based program learners (synthesizers)
generate several programs, rank these programs from most-likely to least-likely
(usually, based on a domain-specific ranking function), and return the best ranked program.
For example, the program learner may return the
programs depicted in Figure~\ref{fig:sample_programs}.

\begin{figure}[t]
    \scriptsize{
        \begin{tabular}{|l|l|}
            \hline
            Input & Desired output \\
            \hline \hline
            12 units PID 24122 Laptop & PID 24122 \\ \hline
            43 units PID 98311 Wireless keyboard & PID 98311 \\ \hline
            7 units PID 21312 Memory card & PID 21312 \\ \hline
            22 units PID 23342 Docking station v2 & PID 23342 \\ \hline
            2 units PID 24232 Mouse with pad & PID 24232 \\ \hline
            1.5 Litres PID 32465-X Printer ink & PID 32465-X
            \\ \hline
            \dots & \dots
            \\\hline
        \end{tabular}
    }
    \caption{Task: Extracting
    product identifiers}
    \label{fig:illustrative_example}
\end{figure}

\begin{figure}[h]
  \scriptsize{
    \begin{tabular}{|l|l|r|}
      \hline
      $P_i$ & Program description & Rank \\
      \hline \hline
      $P_1$ & String from $1^{st}$ upper case to last digit (inclusive at both ends) & 1 \\
      $P_2$ & String from $1^{st}$ `P' to last digit & 2  \\
      $P_3$ & String from $1^{st}$ upper case upto following $2^{nd}$ non-alphanumeric char & 3  \\
            & \multicolumn{1}{|c|}{$\cdots$} &    \\
      $P_{97}$ & String from $1^{st}$ upper case letter upto following $2^{nd}$ space & 97  \\
      $P_{98}$ & String from $10^{th}$ to $18^{th}$ character & 98 \\
      $P_{99}$ & String between $10^{th}$ character and $4^{th}$ occurrence of the digit $2$ & 99 \\
      \hline
    \end{tabular}
  }
  \caption{Sample programs for extracting product identifiers}
  \label{fig:sample_programs}
\end{figure}
Let us assume that the learner ranks $P_1$, $P_2$, and
$P_3$ high (\emph{e.g.} with ranks $1$, $2$, and $3$, respectively); and $P_{97}$, $P_{98}$ and
$P_{99}$ low (\emph{e.g.} with ranks $97$, $98$, and $99$, respectively).
The most-likely program (here $P_1$),
is then run on all the other input strings, producing \emph{likely output strings}.

At this point, the user is faced with an enormous task. The user has to look through
a large set of likely output strings and verify that they are all correct, and if not,
then pick an input where the output is incorrect. 
This input, called a {\em{significant input}}, is used
to provide the second example to the learner, and the process repeats.

We would like to convert the passive program learner into an active learner that suggests a next
{\em{significant input}} in each iteration. The goal is to not only reduce the cognitive load on the user,
but also make the process converge in fewer iterations.

\ignore{
Since it is not feasible to work with the complete set of possible programs
(which is large and potentially infinite) or the full set of inputs (potentially
too large), we need to {\em{sample the space of possible programs}} and
{\em{sample the inputs}}.

Intuitively, our algorithm works as follows:
\begin{inparaenum}[(a)]
\item Sample a set of programs consistent with the given examples;
\item Sample a set of inputs from the given inputs;
\item Find an input from the sampled inputs that distinguishes at least two
  programs from the sampled programs.
\end{inparaenum}
}


It is clear that to change a passive program learner to an active learner,
we need to solve the significant-input generation problem, which in turn is an instance 
of the more-general significant questions problem. 
Hence, we know that a promising
choice for the next significant input should be the input about which 
we have the least amount of knowledge (of the output for that input).
However, computing the entropy of a given input (question) is not trivial.
We need two things. 
First, we need the program learner to synthesize not one program, but a 
probability distribution over the set of all
programs consistent with the given input-output examples.
Second, we need to use that probability distribution to compute the 
entropy $En(i)$ of each input (question) $i$.
Since the number of programs and inputs can be prohibitively large, we need
to {\em{carefully sample}} from these sets to find the input with greatest
uncertainty as the next significant input.



\paragraph{Random Programs, Top Programs, and Combinations.}
We use sampling techniques~\cite{sampling} to estimate entropies of inputs.
One approach for sampling programs 
is to randomly pick programs from the set of all consistent programs.
However, this strategy will likely pick only the low ranked programs since there are many more such 
programs.
If that happens, then the chosen significant input will distinguish low ranked programs,
but not necessarily the currently picked program from (some other high ranked) correct program.
This is a {\em{false positive}}: when presented to the user, the input does not appear as distinguishing 
to the user.
For example, the input $i_5$, ``2 units PID 24232 Mouse with pad'', produces different outputs when given to 
$P_{97}$ and $P_{98}$, and hence  and could be picked as an significant input, 
but if $P_{97}$ was the current candidate program, the user would observe that
its predicted output on $i_5$ actually matches the desired output.

One obvious solution is to consider only highly ranked programs.
However, there are plenty of very similar programs at the top, and these programs tend 
to behave similarly on the given inputs.
For example, $P_1$, $P_2$, and $P_3$ on
the input $i_6 = \text{``1.5 Litres PID 32465-X Printer ink''}$
all produce the output ``PID 32465-X'', and thus $i_6$ does not distinguish them.
Hence, considering only highly ranked programs would quickly cause us to conclude non-existence of significant
inputs, thus increasing the chance of {\em{false negatives}}.
However, $i_6$ is likely to actually be significant because it contains a product ID in a different format.

Aiming to reduce the probability of both false positives and false negatives,
our algorithm samples the program space by including some highly ranked programs as well as
some randomly sampled programs. 
For our example, let us assume $P_1$, $P_2$  are chosen into the sample as the
highly ranked programs, and $P_{97}$ is picked as the randomly-chosen program.
In this case, the input $i_6$ is presented as a significant input as it
distinguishes $P_{97}$ and $P_1$.

\paragraph{Clustering Inputs based on Input and Output Features.}
Sampling inputs is challenging because we need to ensure that potentially significant
inputs are picked in the sample.
Consider, for example, a hypothetical input $i_7 = \mbox{``pid 12345''}$ that is available
in our large universe of possible inputs.
This input is clearly different from the rest, and hence, ideally, it should be picked in our
sample set.
We achieve this goal by using the concept of input clustering: we group the set
of inputs into clusters, where each cluster (described by a regular expression)
contains inputs of a certain ``shape'', and then we sample at least one input from
each cluster.
The input set $\{i_1,\ldots,i_7\}$ is naturally partitioned into two clusters described by the regular expressions
$\mathtt{\alpha^* \mbox{``PID\ ''} [0-9]^5 \alpha^*}$ and $\mathtt{\mbox{``pid\ 12345''}}$.
Hence, $i_7$ is sampled from the second cluster and included in the sample input set.

Just picking $i_7$ in the sample set is not enough. Note that
all the $6$ programs listed above return the same default null value, say $\epsilon$,
on this input, and thus we have ``complete knowledge'' of the output on $i_7$, and
hence it will not be picked as a significant input.
However, intuitively, it is clear that the input $i_7$ is significant to the task.
We solve this challenge by giving special treatment to the default value $\epsilon$.
Specifically, different instances of the value $\epsilon$ are treated
as being {\em{different}} from each other. With this change, we now have almost no
knowledge of the output on the input $i_7$ because all programs produce different outputs on $i_7$.
This causes $i_7$ to turn into a likely significant input.

%

There are certain disadvantages to input clustering. Consider the task of extracting the
area code (first $3$ digits) from phone numbers. A $10$-digit phone number can be written in many
different ways\footnote{For instance, ``1234567890'', ``(123)4567890'', ``123-456-7890'', ``123 456 7890''.}, and hence input clustering will generate lots of clusters. Now, suppose that our algorithm
has successfully  found the desired program, but does not yet know that it has converged.  
There might still exist
inputs that distinguish some programs in the sample program set. However, if we look at the
outputs, 
we notice that the outputs look uniform,
and in fact, clustering the outputs results in just $1$ cluster, which is used as an
indication that the synthesis process has converged. This helps us avoid false positives.
Hence, we sample inputs based on features in both the inputs and the outputs.

\endignore}



%
%

\subsection{From Passive to Active Synthesis}
Let $I, O$ be sets that denote the domain for the input space and the output space respectively.
Let $f: I \mapsto O$ be a fixed function (that is unknown to the program learner, but is
known to the user).
Let $\Sigma := I \times O$ be the set of all possible input-output examples, and
let $\Sigma^*$ denote the set of all finite sequences of these examples.
Let $PS$ be the space of all programs (considered by the program synthesizer) that map $I$ to $O$.
Note that $f$, and every $p\in PS$, maps $I$ to $O$, but the difference is that elements in $PS$ are
computable (executable) descriptions of functions, whereas $f$ is modeling the user.

A {\em{passive program learner}} $ppl$ is a computable function
with the signature
$ppl: \Sigma^* \times 2^{I} \mapsto PS$ such that
for any
input-output example sequence $seq$,
\begin{eqnarray}
seq := \left[ \la in_1, f(in_1)\ra , \ldots, \la in_k, f(in_k)\ra \right], 
    \label{eqn:seq}
\end{eqnarray}
and a subset $I_0 \subseteq I$ of the input space,
the passive program learner 
returns a program $p := ppl(seq, I_0)$ that
satisfies all the given input-output examples; that is,
 $$p(in_j) = f(in_j) \qquad \mbox{ for every } j = 1,2,\ldots,k.$$
The goal of the passive learner is to find a program $p$ that matches $f$ on all inputs in $I_0$, and not
just the inputs in the provided examples.

Existing programming-by-example (PBE) systems can be viewed as passive program learners.
They maintain a sequence $seq$ of input-output examples, which initially is either empty or
contains just one input-output pair. They then generate the program
$p := ppl(seq, I_0)$, and ask the user if the outputs $p(I_0)$ match the expected outputs.
If not, the user provides a new input-output pair that gets added to $seq$ and the process repeats.

Finally, we note that the program $p$ returned by $ppl$ is not arbitrary, but one that is ranked
highest. The ranker is designed to prefer programs that are most-likely to be the user-intended program.
Designing such rankers is not easy: it is often achieved by a combination of machine learning and
human tweaking of ranking function parameters based on user feedback.

\algnewcommand{\LineComment}[1]{\State \(\triangleright\) #1}
\begin{figure}[t]
    \centering
    \begin{algorithmic} 
        \Require $ppl^*$, a modified passive program learner
        \Require $I_0$, a subset of inputs
        \Require $\epsilon > 0$, an uncertainty threshold
        \Function{ActiveProgramLearner}{$ppl^*, I_0$}
            \State Input-output examples $\spec \gets [\;]$
            \State Prob. Dist. $pd \gets \mbox{domain-dependent prior on $PS$}$
            \State Entropy (uncertainty) about desired program $un \gets En(pd)$
            \While{$un \geq \epsilon$} \Comment{while there is some uncertainty about intended program}
                \State foreach $i\in I_0$: $Pr_i \gets \lambda{a}: \sum_{p\in PS, p(i)=a} pd(p)$
                \State \Comment{$Pr_i(a)$ is probability of $a$ being output on $i$}
                \State $i \gets \argmax_{i\in I_0} En( Pr_i )$
                \Comment{input with greatest uncertainty}
                \State $\spec \gets \spec \sqcup \la i,\, f(i) \ra$
                \Comment{Call oracle $f$ to get $f(i)$ and add $(i,f(i))$ to $\spec$}
                \State  $pd \gets ppl^*(\spec, I_0)$
                \Comment{Update belief prob. dist. about intended program}
                \State $un \gets En(pd)$
                \Comment{Update uncertainty about intended program}
            \EndWhile
            \State \Return $p_{best} = \argmax_{p\in PS} pd(p)$
        \EndFunction
    \end{algorithmic}
        \caption{An active program learner that selects the next significant input (to query the oracle/user) at each iteration greedily based on information gain.}\label{fig:integratedProc}
\end{figure}

\subsubsection{Active program learner.}
We now turn the passive learner into an active learner.
        Procedure~$\texttt{ActiveProgramLearner}$ in Figure~\ref{fig:integratedProc} uses greedy information
        gain (Proposition~\ref{prop:main}) to 
        implement an active program learning method.  The procedure maintains its current belief of the
        intended program as a probability distribution $pd$ on the program space $PS$.
        The entropy, $En(pd)$ of this distribution is a measure of our uncertainty, and while 
        our uncertainty measure is greater-than a threshold $\epsilon$, we continue to add an
        input-output $\langle i, f(i)\rangle$ to the set $\spec$ of input-output examples.
        The input $i$, which is picked in each iteration as a significant input, is the one that
        maximizes entropy $En(Pr_i)$.  Note that
        $En(Pr_i)$ is the uncertainty in the output for input $i$ given our current belief
        $pd$ of the desired program.  
        Once a new input-output pair is added to $\spec$, we use an enhanced passive learner, $ppl^*$,
        to update $pd$ in that iteration. 
        When the loop terminates, we return the program $p$ whose probability
        $pd(p)$ is maximum as the learnt program.

        The active program learner uses an {\em{enhanced}} passive learner, $ppl^*$, as a subroutine.
        The key difference between $pp$ and $ppl^*$ is that $ppl^*$ returns a probability distribution $pd$
        on the program space, and not just a single program.
        Since computing and representing $pd$ precisely is not feasible,
        the probability distribution $pd$ is returned in the form of a sampled set of 
        programs (consistent with the input-output examples generated so far) and an assignment of
        probability to this sampled subset.

        \ignore{ 
\paragraph{Convergence Criteria.}
We evaluate active program learners in the following way. We enclose the while loop in 
        Procedure \texttt{ActiveProgramLearner}
        inside an outer loop that terminates only when $p_{best}(i) = f(i)$ for all $i\in I_0$,
        where $p_{best}$ is the program computed at termination of the while loop.
        If the inner loop terminates, but the outer does not, then we pick an input $i$
        where $p_{best}(i) \neq f(i)$ as the next significant input and continue with the
        procedure.
        Active program learners are evaluated based on the following three measures.

%
\begin{description}
	\item[Number of iterations:] The number of iterations until the outer loop terminates.
	\item[False positives:] 
		Whenever the inner loop generates an input $in$ on which the
                current program $p_{best}$ and the desired function $f$ agree (that is, $p_{best}(in) = f(in)$),
		the resulting iteration {\em{appears}} futile to the user. Such inputs $in$ are called
        false positives, and we want to minimize them.
	\item[False negatives:] 
		When the inner loop terminates, it communicates to the user that
        the current program candidate $p_{best}$ is considered correct and
		there are no further ``significant inputs''.
        However, if $p_{best} \ne f$ on some input, then this conclusion is incorrect, and we have a false negative. We want to minimize the number of false negatives.
\end{description}
The three criteria above differ in importance in different applications.
Generally, false negatives are more expensive than false positives since a false negative requires the user to
manually find the next distinguishing input in $I_0$, whereas a false positive requires only a confirmation of the
current program output.
However, they also differ in their cognitive load and user experience implications: a false positive is likely to cause
irritation and mistrust in the system, whereas a false negative may lead the user toward ending the interaction
prematurely and using an incorrect synthesized program.

\endignore}

\ignore{
\ap{The following 2 paragraphs and the corresponding figure are rather trivial, and can be explained in two sentences in
the Evaluation section.}

To evaluate significant-input generators, we
change the termination condition in Procedure~$\texttt{InteractiveSynthesis}$
	so that we terminate only when the user is convinced
	that the learnt program $p$ matches the desired function $f$ exactly.
	This version of the synthesizer is shown in Procedure~\texttt{InteractiveSynthesizer}
	in Figure~\ref{fig:interactive}.
	Procedure~\texttt{InteractiveSynthesizer} also maintains some additional statistics.
	The variable $\texttt{numFalseNegatives}$ counts the number of times the significant-input generator
	returns $\bot$ before termination.
	The variable $\texttt{numFalsePositives}$ counts the number of times the significant-input generator
	returns an input on which the synthesized program $p$ already computes the correct output.

Ideally, we desire a significant-input generator that ensures that the resulting incremental synthesizer
always terminates
with $\texttt{numFalsePositives} = \texttt{numFalseNegatives} = 0$ and
with minimal value for $\texttt{numIterations}$.

\begin{figure}[t]
\centering
\begin{algorithmic} 
\Function{InteractiveSynthesizer}{$ps, sig, I_0$}
\State $\spec \gets \la \ra$; \qquad $i \gets 0$;
	\State $numFalsePositives \gets 0$; \qquad $numFalseNegatives \gets 0$;
\State $p \gets $ default program
	\While{ $f|_{I_0} \neq p|_{I_0}$}
	\State $in \gets sig(ps, \spec, I_0)$
	\If{ $in == \bot$ }
	  \State $numFalseNegatives \gets numFalseNegatives + 1$
	  \State $in \gets$ randomly picked input on which $f$ and $p$ differ
	\Else
	  \If { $p(in) == f(in)$ }
	    \State $numFalsePositives \gets numFalsePositives + 1$
	   \EndIf
	\EndIf
	\State $\spec \gets \spec \cdot (in, f(in))$; \qquad $i \gets i + 1$;
	\State  $p \gets ps(\spec, I_0)$
	\EndWhile
\State \Return $p, i, numFalsePositives, numFalseNegatives$
\EndFunction
\end{algorithmic}
	\caption{$\texttt{IncrementalSynthsizer}$: Incremental Program Synthesis Using Significant Inputs}\label{fig:interactive}
\end{figure}
}




\ignore{
Given a probability distribution $pd$ on the program space $PS$
and a subset $I_0$ of inputs,
a {\em{significant-input generator}} $sig$ is a procedure that
returns either an input $in_{k+1}\in I_0$ or a special value
$\bot$.
\begin{eqnarray*}
    sig : ({{PS}} \mapsto [0,1]) \times 2^{I} \rightarrow \{\bot\} \cup I
\end{eqnarray*}
Intuitively, a significant-input generator determines if the learning process has converged
(it returns $\bot$ in that case),
and if not, it returns an input $i$ in
$I_0$ such that knowledge of the example $(i, f(i))$ will help expedite convergence.
\endignore}

	\subsection{Sampling Program Learner}
\def\ioexs{{\spec}}
The enhanced passive program learner, $ppl^*$, is implemented as a sampling program learner.
A {\em{sampling program learner}} $spl$ is a computable function
with the signature
$spl: \Sigma^* \times 2^{I} \times \SamplingSpec \mapsto (PS \mapsto [0,1])$ such that
for any
input-output example sequence $seq$, subset $I_0$ of inputs, and a sampling specification $sspec$, 
the returned probability distribution $pd := spl(seq, I_0, sspec)$ is such that
\\
(a) if $pd(p) > 0$ then $p$ is consistent with all examples in $seq$, 
\\
(b) the set $\{p \mid pd(p) > 0\}$ is consistent with the sampling specification $sspec$.
\\
The probability distribution returned by $spl$ is assumed to reflect the current belief about the
intended program.

The reason for working with samples is obvious: the space of programs consistent with a given set of input-output examples can be very large.
It has been observed that a typical real-life domain-specific language for data transformation may contain up
to $10^{20}$ programs consistent with a given single input-output example~\cite{cav:ranking}.
While such a program space can be represented symbolically using version space
algebras~\cite{vsa:Mitchell,lau:smartedit,flashmeta} or finite tree automata~\cite{wang17,wang18},
working with a probability distribution over this space is infeasible and counterproductive.
We therefore work with samples. 

A {\em{sampling specification}} $(top,random)$ is a pair of numbers that indicate how
many top programs to pick and how many random programs to sample from the set of
programs consistent with the given input-output examples.
The procedure for collecting $k$ top programs poses no significant challenges: most passive
program learners that can generate one top program can also generate $k$ top programs.
So, we just focus on random sampling here.

First, we observe that a sampling specification is trivial to satisfy when
the set of all programs consistent with the correctness specification (input-output examples)
is small. We also note that we can sample from
state-of-the-art symbolic program set representations, including VSAs~\cite{lau:smartedit} and
FSAs~\cite{wang18}.\footnote{This is more challenging when an underlying program synthesizer is based on a constraint
    solver, where requesting top $k$ programs is feasible for state-of-the-art optimizing solvers~\cite{bjorner2015nuz}
but requesting \emph{uniformly random} instances is hard.}
We next describe how to randomly sample {\em{while performing synthesis using an enumerative~\cite{transit} or deductive approach~\cite{flashmeta}}}.

\begin{figure*}[t]
    \begin{mathpar}
        \text{\bfseries R1} \qquad
        \inferrule
            {\mathit{Synth}(N,\phi) = \mathit{Synth}(f(N_1,N_2),\phi) \cup \mathit{Synth}(g(N_3,N_4),\phi)}
            {\mathit{RandomK}(N,\phi) = \mathit{UniformK}( \mathit{RandomK}(f(N_1,N_2),\phi) \cup \mathit{RandomK}(g(N_3,N_4),\phi) )}
        \\
        \text{\bfseries R2} \qquad
        \inferrule
            {\mathit{Synth}(f(N_1,N_2),\phi) = f( \bigcup\nolimits_i \mathit{Synth}(N_1,\phi_{1i}), \bigcup\nolimits_j \mathit{Synth}(N_2,\phi_{2j})}
            { \mathit{RandomK}(f(N_1,N_2),\phi) = \mathit{UniformK}( f(\bigcup\nolimits_i \mathit{RandomK}(N_1,\phi_{1i}),
                \bigcup\nolimits_j \mathit{RandomK}(N_2,\phi_{2j})) ) }
    \end{mathpar}
    \caption{{\small{Sampling program learner extends the baseline program learner with sampling specification that are propagated down to subproblems, and then the results combined. Here $\mathit{UniformK}$ denotes a uniform sampling of the set in its argument.}}}\label{fig:samplinglearner}
\end{figure*}

Consider the program space $PS$ that consists of programs generated by a top-down tree grammar.
Let $N := f(N_1,N_2) \; | \; g(N_3,N_4)$ be two top-down tree automata transitions
that say that a program generated by nonterminal $N$ can either be of the form
$f(p_1,p_2)$, or of the form $g(p_3,p_4)$, where each $p_i$ is
recursively generated from nonterminal $N_i$.
While there are many approaches for program synthesis, the preferred inductive synthesis
approach is based on decomposing a synthesis problem,
$\mathit{Synth}(N,\phi)$, on a nonterminal $N$ and a specification
(input-output examples) $\phi$ into subproblems on nonterminals $N_1,\ldots,N_4$ and
derived specifications $\phi_1,\ldots,\phi_4$, and subsequently, putting the results of
the subproblems together to obtain a solution for the original problem.
The key idea behind a {\em{sampling program learner}} is that we can extend this decomposition
step to {\em{also decompose the sampling specification $ss$}}.
This means that we
decompose the learning problem $(N,\phi,ss)$, where $N$ is the nonterminal,
$\phi$ is a program correctness specification, and $ss$ is a sampling specification,
into subprogram learning problems,
$(N_i, \phi_i, ss_i)$, for $i=1,\ldots,4$,
and after we have
recursively solved the subproblems, we obtain a solution for the original
problem by composing the solutions together.
In particular, for sampling, this means we get samples of subprograms, and
we use them to get samples for the top-level program.

Figure~\ref{fig:samplinglearner} recursively defines the function
$\mathit{RandomK}(N,\phi)$, which returns
$k$ random samples of programs generated by $N$ and consistent with $\phi$.
Its definition follows the definition of the passive learner, $\mathit{Synth}$, itself. In particular,
\begin{description}\itemsep=0em
    \item[(R1)]
if $\mathit{Synth}$ decomposes the synthesis problem on $N$ and $\phi$
to synthesis over $f(N_1,N_2)$ and $g(N_3,N_4)$, then
we uniformly sample from a set containing $k$ random samples of the form
$f(N_1,N_2)$ and $k$ of the form $g(N_3,N_4)$, and
    \item[(R2)]
if $\mathit{Synth}$ decomposes the synthesis problem on $f(N_1,N_2)$ and $\phi$
in terms of subproblems on $N_1$ and $N_2$, and gets its result in the form
$f( \bigcup_i P_i, \bigcup_j P_j)$, then $\mathit{RandomK}$ samples equal number from each $P_i$ to get
the $k$ random samples of $f$-rooted programs.
\end{description}
Note that random sampling is not uniform over the program set, but uniform over the 
syntactic classes of programs that are generated during the synthesis process: this is ideal because it ensures that samples are diverse.

We omit several low-level details about sampling program learners
here, e.g. sub-specifications can be conditioned on other
sub-specifications. However, most of these details are easy to extend to {\em{sampling}}
specifications by following the approach taken for {\em{correctness}} specifications~\cite{flashmeta}.

\ignore{ 
Let $\ioexs = \left[ \la in_1,out_1 \ra, \ldots, \la in_k,out_k \ra\right]$ be a sequence of
$k$ input-output examples.
If $P$ is the universe of programs (that is being considered by the synthesizer),
then the set of programs consistent with the examples in \spec is defined as
$$
    P|_{\ioexs} := \{ p\in P \;\mid \; p(in_i) = out_i \text{ for } i = 1 \dots k \}
$$
The set $P|_{\ioexs}$ represents all the ambiguous programs that may solve the intended user task.
As mentioned before, it can be very large and infeasible to enumerate directly.
Furthermore, any sampling strategy over $P|_{\ioexs}$ must balance a trade-off between prioritizing high-ranked programs
(to prefer programs with a priori good generalization) and prioritizing sample diversity (to maximize the likelihood of
the desired task-specific program in the sample).

\begin{example}[Large program sets]
    Consider the scenario where the user wants to concatenate several tokens
    where each token can be extracted from the input in many different ways.
    For instance, let the first input-output example is $i_1\mapsto o_1$ where
    $i_1$ is ``123 ABC 456 DEF 789''  and $o_1$ is ``123456789''.
    Let us say there are
    $n_1$ highly and closely ranked programs that extract the substring ``123'' from $i_1$,
    $n_2$ such programs that extract the substring ``456'' from $i_1$, and
    $n_3$ such programs that extract the substring ``789'' from $i_1$.
    Then, all the top programs that transform $i_1$ to $o_1$ will consist of
    the $n_1 \cdot n_2 \cdot n_3$ combinations of these subprograms.
    A sample of high-ranked programs from this combination will only contain a limited number of different subprograms.
    Thus, a strategy that maximizes sample diversity should also include lower-ranked subprograms sampled at every level
    of the synthesis process.
\end{example}

The considerations above suggest the following requirements on a program sampling strategy and the underlying program
synthesis engine as part of a significant-input generator:
\begin{itemize}\itemsep=0em
    \item Given a ranking function $h\colon P \to \mathbb{R}$ that optimizes program generalization, the strategy should
        be able to request top $k$ programs w.r.t. $h$ from the set~$P|_{\ioexs}$, for any fixed $k$.
    \item Given a set \spec of input-output examples, it should be possible to construct a uniform sample from the set
        of consistent programs $P|_{\ioexs}$.
\end{itemize}

\endignore}

\paragraph{Probability measure function.} Finally, we need to assign probabilities to the sampled programs.
We assign a probability to a sampled program that is proportional to its rank order. 
Note that our sample contains some top-k program and some randomly sampled programs.
All (passive) program learners are equipped with a ranking function
that assigns a rank to each (synthesized) program; however, these ranks do not directly map to probabilities
in any way, but are only coarse indicators of a program's likelihood to be the intended program.
The top-$k$ programs are picked based on this ranking.
While the rank (score) itself is not meaningful, the {\em{order}} it induces on the programs remains
meaningful, and hence assigning a (slightly higher) probability to samples
that are ranked higher is justified.

In \Cref{sec:evaluation}, we show the value of the combination of top-$k$ and random sampling by
performing evaluation using
three specific approaches: (a) no sampling, (b) only top-$k$ programs, and
(c) a combination of top-$k$ and uniform sample.


	\subsection{Input Sampling}
Input sampling is required because real-life datasets in data wrangling scenarios typically contain tens of thousands of
rows, and enumerating them all is counterproductive for multiple reasons.
First, a typical UI response time for a user-facing application must stay within 0.5 sec, which is difficult to satisfy
when enumerating over the entire dataset.
Second, most inputs in a typical dataset have similar \emph{distinguishability}: ideally we should consider just one
representative.

\begin{table}[t]
    \begin{center}
    \begin{tabular}{c||c|c|c||c}
        & \multicolumn{3}{c||}{Programs} & Entropy
        \\ \hline
         Inputs & $p_1$ & $p_2$ & $p_3$ & $En(i)$ 
        \\ \hline
        $i_1$ = ``foo1bar11baz'' & 1 & 1 & 11 & $\sum_{i=1}^{2}-\frac{i}{3}\log\frac{i}{3}$ 
        \\ 
        $i_2$ = ``foo2bar22baz'' & 2 & 2 & 22 & = 0.9 
        \\ \hline
        $i_{95}$ = ``fooabara1baz'' & \mbox{a} & \mbox{a} & \mbox{a1} & = 0.9 
        \\ \hline
        $i_{96}$ = ``fooabar-1baz'' & \mbox{a} & \mbox{-} & \mbox{-1} & $\log(3)$= 1.6 
        \\ \hline
        $i_{97}$ = ``uvw'' & $\epsilon$ & $\epsilon$ & $\epsilon$ & {\bf{1.6}} 
    \end{tabular}
    \end{center}
    \caption{{\small{Clustering inputs: Assume that the three programs $p_1,p_2,p_3$
    have equal probability ($\frac{1}{3}$), and that they produced the shown outputs on the five inputs. Based on features in the input strings, Inputs $i_1$ and $i_2$ are similar, and even have equal entropy $0.91$, and should be clustered.
    Inputs $i_{95},i_{96}$ may
    get clustered with $i_1$,$i_2$ based on input features, but based on features in the output (columns
    $o_1$, $p_2$, and$p_3$)
    they are likely to be in different clusters, which is good since $i_{96}$ has higher entropy.
    }}}\label{table:clustering}
\end{table}


\def\ninputs{{n}}
\def\partition{{\texttt{Partition}}}

Input sampling aims to reduce the computational cost of the active program learner by
restricting it to a sample of the input space $I$.
The na\"ive uniform sampling approach is not ideal here,
and 
we use two key ideas for sampling inputs:
(1) input-features based clustering and (2) output-features based clustering.

\paragraph{Input-features based clustering.}
The hypothesis here is that {\em{reasonable programs behave similarly on inputs with similar features,
and hence, such inputs are likely to have the same uncertainty}}, as illustrated in Table~\ref{table:clustering}.
Hence, we cluster the inputs based on
{\em{string clustering}}, and sample equally from each cluster to ensure
full coverage of different ``shapes''.

A string clustering algorithm takes a dataset $I$, and returns a \emph{partition} of this set into
disjoint clusters.
It is parameterized with a similarity measure to cluster the strings in $I$.
Formally, it has the following signature:
$$
\texttt{cluster} : I  \mapsto  \left(I \mapsto \{1,2,\ldots,M\}\right)
$$
where $M$ is the (maximum) number of clusters created.
A {\em{partition}} is a function $\partition: I \mapsto \{1,\ldots,M\}$ that
maps each input $in$ to one of the $M$ clusters.
Let $I^i = \{ in \;\mid\; \partition(in) = i \}$ denote the $i$-th cluster.

Intuitively, inputs in different clusters should have sufficiently different syntactic shape.
Thus, a \emph{diverse} uniform sample $I^*$ of $\ninputs  = |I|$ inputs can be constructed
by randomly sampling $\lceil\ninputs * |I^i|/|I|\rceil$ inputs from the $i$-th cluster $I^i$.

Clustering is parameterized by a similarity measure on the input space,
which is based on standard features extracted from strings; see~\cite{matching-text} for details.

\begin{example}
    Consider the data transformation task where a user has presented one example
    ``12\ in'' $\mapsto$ ``12'', and the set of other inputs includes ``8\ in'' and ``30\ cm'' (and other
    strings denoting length in either $in$ or in $cm$).
    Most candidate programs that are learnt from the one given example are unlikely to perform differently
    on ``30\ cm'', and hence it is unlikely that ``30\ cm'' will be a distinguishing input.
    However, input clustering will clearly identify two separate clusters corresponding to the two units,
    and  ``30\ cm'' should be presented as a significant input to the user.
\end{example}

\paragraph{Output-feature based clustering.}
The hypothesis here is that {\em{if the output on input $i$ looks sufficiently different from the outputs generated by other inputs, then uncertainty about $i$ is likely to be high.}}
Hence, the outputs generated by the current candidate programs can indicate which inputs are potential candidates for being significant.
Table~\ref{table:clustering} shows a $2$-d matrix over programs and inputs:
output-based clustering partitions inputs based on clustering the values in Column~$p_1$ 
(where $p_1$ is the top-ranked program). We can optionally also cluster based on Column~$p_2$ and Column~$p_3$.
Whereas the entropy $En(i)$ of an input $i$ is defined by the values in $i$'s row, 
$En(i)$ having a low value often correlates with $p_1(i)$ being of a different ``shape'' than other
outputs in Column $p_1$ (see Input $i_{95}$ and $i_{96}$).


\begin{example}
    Consider a scenario where the user is extracting the year from dates formatted
    in many different forms.
    The given input set contains the inputs
    ``05-Feb-2015'',
    ``25 December 2013'',
    ``2010-12-12'',  and
    ``9/3/2017''.
    Clustering the input space gives a large number of partitions.
    However, the output set generated by any synthesized candidate program should be relatively uniform (in this
    scenario, form a single cluster described as \texttt{\textbackslash d\{4\}}).
\end{example}

\paragraph{Null outputs.}
Exceptions need to be handled properly when computing the uncertainty $En(i)$ about an input $i$.
Specifically, when an input does not satisfy the preconditions of
a (synthesized) program, its output defaults to a special value null value, denoted as, say, $\epsilon$.
However, not all such values are identical, and hence, when defining uncertainty about an input,
we treat every instance of a null value in the output as being different from each other.
\begin{example} 
  String transformation programs often return a null value when the input is
  not of the format they expect.
  For example, the programs ``extract the first digit'' and ``extract the second
  digit'' both return the null value on the string ``ABC''.
  However, it is a stretch to say that these programs behave similarly on the
  input ``ABC''. Therefore,
  we consider all null values to be unequal to each other when defining
    the uncertainty, $En(i)$, about (the output on) input $i$. In Table~\ref{table:clustering},
    Input~$i_{97}$ generates null values on all programs, and hence, with this change,
    $En(i_{97})$ is not $0$ but the larger value $1.6$.
\end{example}

        \subsection{Information Gain and Distinguishability}
\def\DistPP{{\mathtt{DistPP}}}
\def\ZeroOne{{\mathtt{ZeroOne}}}
\def\meas{{\textsf{dm}}}
\def\entropy{{\textsf{ig}}}
\def\prob{{\mathbb{P}}}

The uncertainty $En(i)$ about an input $i$ is closely related to its ability to distinguish programs in $PS_1$:
recall that an input $i$ distinguishes programs $p_1$ and $p_2$ if $p_1(i)\neq p_2(i)$~\cite{bitvectors}.
The following proposition states that optimizing for uncertainty is at least as general as 
optimizing for distinguishability.
\begin{proposition}\label{prop:measure}
  Given two inputs $in_1$ and $in_2$, if $p_i(in_1) \neq
  p_j(in_1) \implies p_i(in_2) \neq p_j(in_2)$ for all programs $p_i,p_j$, then 
    $En(in_1) \leq En(in_2)$.
\end{proposition}

\ignore{
\paragraph{The Weighted Cardinality Measure.}
\emph{Principle~1} hints that the cardinality of $P^*(in)$ is proxy for the goodness
of a significant input $in$. 
However, this does not account for \emph{Principle~2}.
The weighted cardinality measure is an extension of cardinality to account for
the ranks of the programs that produce $P^*(in)$.

Let $P^* = \{ p_1, p_2, \ldots , p_n \}$ where the synthesizer ranks $p_i$
higher than $p_j$ for all $i < j$.
We use a vector characterization $\overline{P^*(in)}$ of $P^*(in)$ that
shows the first position of each distinct output in $P^*(in)$.
Formally, $\overline{P^*(in)}$ is an $n$-dimensional zero-one vector where the
$i^{th}$ component is $1$ iff $\forall j < i : p_i(in) \neq p_j(in)$.

The \emph{weighted cardinality measure} $\meas_W$ is parametrized by a weight
vector $\overline{W} = \left(W_1, W_2, \ldots, W_n\right)$, where each $W_i$ is
positive. 
We define $\meas_W(in, P^*)$ as $\langle \overline{W} , \overline{P^*(in)}
\rangle$, where $\langle \cdot , \cdot \rangle$ denotes the standard inner
product.
Note that if each $W_i$ is $1$, the measure is equal to the cardinality of the
set $P^*(in)$.
\endignore}

\ignore{

We show a form of optimality $\meas_W$: it is possible to define $\overline{W}$
such that the expected sum of ranks of the programs eliminated by knowing the
true (user intended) output for a significant input $in$ is minimized.
In $P^*$, assume that the probability that the program $p\in P^*$ is the desired
program is $pr(p)$.
The probability $pr(p)$ could potentially be generated from the ranking measure
used by the program synthesis engine.
Define the importance index of program $p_i$ as $n-i+1$, so that the best candidate program
$p_1$ has the highest importance index.
Define the weight vector $\overline{W}$ such that $W_i = (n-i+1) * (1-
\sum_{p(in) = p_i(in)} pr(p))$.

\begin{proposition}\label{prop:weights}
  Given the ranked program set $P^* = \{p_1,\ldots,p_l\}$, probabilities
  $pr(p_i)$ that each $p_i$ is the desired program, the weight vector
  $\overline{W}$ defined above:
  then the input $in \in I$ with the highest measure $\meas_W(in)$  is the input
  that maximizes the expected importance index of the most important program to
  be likely eliminated if $in$ is chosen as the significant input.
\end{proposition}

Thus, by picking suitable weight matrices $W$ we can define different measures that
optimize different criteria.
In our experiments, for each $P^* = \{ p_1, \ldots, p_n \}$, we use the weight
vector $(n, n-1, \ldots, 1)$.
However, other, more sophisticated weight vectors (\emph{e.g.}, based on
discounted cumulative gain) can be considered.

\endignore}


\subsection{Evaluation}
\label{sec:evaluation}
We evaluate active program learners in the following way. We enclose the while loop in
        Procedure \texttt{ActiveProgramLearner}
        inside an outer loop that terminates only when
        the program $p_{best}$ learnt by inner loop is consistent with {\em{all}} the input-output examples;
        that is, $p_{best}(i) = f(i)$ for all $i\in I_0$.
        The outer loop is intended to mimick interaction with the user, which is needed
        whenever the inner loop terminates, but the outer does not. These cases are counted
        as {\em{false negatives}}, and in such cases, we continue the inner loop by picking an input $i$
        where $p_{best}(i) \neq f(i)$ as the next significant input.
        Active program learners are evaluated based on: 
%
%
\begin{description}
	\item[Number of iterations:]
        We want to minimize the number of iterations of the inner loop until the outer loop terminates.
	\item[False positives:]
		Whenever the inner loop generates an input $in$ on which the
                current program $p_{best}$ and the desired function $f$ agree (that is, $p_{best}(in) = f(in)$),
		the resulting iteration {\em{appears}} futile to the user. Such inputs $in$ are called
        false positives, and we want to minimize them.
	\item[False negatives:]
		False negatives occur when the active learner terminates with an unintended program.
        We want to minimize the number of false negatives.
\end{description}

The three criteria above differ in importance in different applications.
Generally, false negatives are more expensive than false positives since a false negative requires the user to
manually find the next distinguishing input in $I_0$, whereas a false positive requires only a confirmation of the
current program output.
However, they also differ in their cognitive load and user experience implications: a false positive is likely to cause
irritation and mistrust in the system, whereas a false negative may lead the user toward ending the interaction
prematurely and using an incorrectly synthesized program.


We evaluated $9$ different variants of our active program learner
on a collection of \numberOfBenchmarks scenarios\footnote{Part of the benchmarks
were taken from \url{https://github.com/Microsoft/prose-benchmarks.}}.
The goal of the variants is to showcase the value of each
key idea proposed in this work.
These variants are defined by their choice in the
(a) program sampling (PS) dimension (``top-$k$'', and ``top-$k$ $\cup$ random''), and the
(b) input sampling (IS) dimension (random sampling, input clustering, output clustering, input+output clustering).
Apart from the $8$ variants obtained from the above choices, we had one baseline version.

The baseline is a {\em{passive program learner}} where we do not use
information gain to pick significant inputs, and let the user do the job. That is,
the implementation just picks the first input where the output of the current program does not match the intent
(mimicking what the user would have to do when interacting with a passive learner).
Thus, in this baseline, the number of false positives is $0$, but
every iteration adds $1$ to the number of false negatives.
The goal of the (eight variants of the) active learner is  to reduce the number of false negatives (the most important criterion) by
potentially increasing the number of false positives.

We remark that the baseline is a state-of-the-art and not naive: the input that the user picks to provide an example
is, in fact, a {\em{distinguishing input}}~\cite{bitvectors,godefroid2012automated}.
One can argue that a smart user might pick a more informative input, but this paper shows that the 
active learner can actually mimick such smart users, and thus reduce the cognitive load on such users.

Procedure~$\texttt{ActiveProgramLearner}$ uses a threshold $\epsilon$ and compares it to the
entropy, $En(pd)$, of the probability distribution over possible programs, $pd$, to decide when to terminate.
Since computing $En(pd)$ just for this purpose is wasteful,
in our implementation, the active learner terminates the session when
none of the top-ranked programs are distinguished by the (maximum entropy) significant input.
%

\begin{table*}[t]
    \begin{tabular}{l|l||c|c|c|c|c||c||c||c}
        & Significant Input& \multicolumn{5}{c||}{\#Iterations} &     \multicolumn{1}{c||}{\#False}  &  \multicolumn{1}{c||}{\#False} & \#Time-
        \\
        & Algorithm Variant & $\leq 1$ & $\leq 2$ & $\leq 3$ & $\leq 4$ & $\leq 32$ & Positives & Negatives & Outs
        \\ \hline
        & Baseline &
        {\bf{11}} & {{39}} & {{109}} & {\bf{184}} & 737 & {\bf{0}} & {{4840}} & {\bf{47}} 
        \\
        & Top-$k$  & 
        {\bf{11}} & {{39}} & 108 & 182 & {\bf{738}} & 5049 & 42 & {\bf{47}}
        \\
    \begin{turn}{90} \makebox[0em][l]{PS}\end{turn}
        & Top-$k$ $\cup$ Random-$k$ & 
        {\bf{11}} & {\bf{41}} & {\bf{110}} & {\bf{184}} & 732 &  5082 & {\bf{3}} & 53
        \\ \hline
        & Random & 11 & 41 & 110 & 184 & 732 & 5082 & {\bf{3}} & 53
        \\
        & Input Clustering & 315 & 498 & 609 & 667 & 733 & 724 & 158 & 52
        \\
        & Output Clustering & {\bf{409}} & {\bf{611}} & {\bf{690}} & {\bf{717}} & {\bf{742}} & {\bf{276}} & 193 & {\bf{43}}
        \\
    \begin{turn}{90} \makebox[0em][l]{IS}\end{turn}
        & Input-Output Clustering & 296 & 481 & 597 & 655 & 733 & 783 & 139 & 52
        \\ \hline
        & Baseline (user)  &
        11 & {{39}} & {{109}} & {{184}} & 737 & {\bf{0}} & {{4840}} & {{47}} 
        \\
        & False Positives & {\bf{409}} & {\bf{611}} & {\bf{690}} & {\bf{717}} & {\bf{742}} & {\bf{276}} & 193 & {\bf{43}}
        \\
    \begin{turn}{90} \makebox[0em][l]{Overall}\end{turn}
        & False Negatives & 296 & 481 & 597 & 655 & 733 & 783 & {\bf{139}} & 52
        \\ \hline
    \end{tabular}
    \caption{{\small{Number of scenarios (out of \numberOfBenchmarks) solved using up to \#Iterations iterations by
    variants of the active programs learner.
        We also show the number of false positives (\#FalsePositives)
        and false negatives (\#FalseNegatives) generated across all scenario instances for these
    variants.
    All runs share a timeout of $60$ sec, with a median time of $0.8$ sec per iteration and mean time of $1.3$
    sec per iteration.
    }}}\label{table:d1}
    \vspace{-1\baselineskip}
\end{table*}

\paragraph{Evaluating program sampling strategies.}
The top part of Table~\ref{table:d1} shows the change in performance of the active program learner as we
change the program sampling technique.
The input sampling technique is fixed to random.

Compared to the baseline, where we have a very high number of false negatives and $0$ false positives,
when using top-$k$ programs as our sample, we increase the number of false positives (because we generate inputs that
distinguish between irrelevant programs), but significantly decrease the number of false negatives.
Combining top-$k$ with random-$k$ gives enough diversity to the program sample to further reduce false negatives, but adds slightly more false positives.
The number of scenarios solved in a given number of iterations does not change substantially.
This experiment clearly shows that top-$k$ \emph{and} random-$k$ programs is the best choice for sampling programs.

\paragraph{Evaluating input sampling strategies.}

The middle part of Table~\ref{table:d1} shows the effect of changing the input sampling technique
on the performance of the active program learner.
We fix program sampling to top-$k$ combined with random-$k$ here.
The ``Random'' sampling strategy is implemented as follows: if the total number of inputs is less-than
a parameter $M$, then it returns all the inputs, and otherwise it samples $M$ inputs randomly from the
set of all inputs.
It turns out that a large percentage of our benchmarks contained a small number of inputs (less-than $M$).
Consequently, random sampling picked the complete
set of inputs, causing very few false negatives.
Clustering causes the active learner to consider only a selected number of {\em{pertinent}} inputs:
this reduces the number of false positives, but since we are ignoring inputs, it adds more false negatives.
Since clustering focuses the active learner on promising inputs, we see a drastic improvement in the number of
benchmarks solved with just 1, 2, or 3 iterations.
Output clustering aggressively removes inputs, and hence, it reduces false positives dramatically, but at the cost
of slightly increasing false negatives.
The results show the value of clustering -- especially when the number of available inputs is really large --
and trade-off between reducing false negatives and false positives.

\paragraph{Overall Evaluation.}
In the bottom part of Table~\ref{table:d1}, we compare the baseline with the version that
optimizes for false positives and the version that optimizes for false negatives (ignoring
the version that use all ``Random'' for input sampling  because they are essentially using
all inputs).
We see that the best active program learners based on greedy information gain
perform much better than what the user is able to achieve
interactively, while also significantly reducing the cognitive load (\#false negatives) on the user.

\section{Related Work}

\paragraph{Query filtering in active learning.}
Since we are not synthesizing inputs, but just picking the ``best'' input to send as a query to the user,
our work falls under the query filtering paradigm of active learning~\cite{CohnAtlasLadner1990}. 
A particular filter, 
called {\em{query by committee}} (QBC)~\cite{SeungOpperSompolinsky1992,FreundSeungShamirTishby1997}, 
works by sampling a committee of (consistent) programs, sampling (randomly) an input (query),
and evaluating entropy of the input on that sample (using our terminology) to either pick or reject it.
This is similar to our work where we sample the programs to evaluate entropies of inputs.
The main difference is that the work on QBC is mostly a theoretical study that makes many assumptions,
such as, existence of a uniform sampling algorithm from the version space. Our work shows how the
same concepts can be applied to a real program synthesis task. Moreover, we also discuss ways to
sample programs and even sample inputs in a way to make the QBC ideas practical in the program
synthesis setting. In the field of program synthesis, the QBC paradigm was used very recently to 
pick queries when synthesizing datalog programs~\cite{Naik2018}. However, it does not formally cast
the program synthesis problem in a probabilistic framework as we do here. Furthermore, the output space
is Boolean (unlike in our setting, where it is String, which causes us to introduce novel ideas, such as,
output clustering), and allows program sampling  to be ``complete'' in a sense (by picking a most-specific
and most-general program from the version space). This is not possible in our more general setting.
This difference also manifests in the fact that 
~\cite{Naik2018} has a complete procedure.

\paragraph{Input and Output Clustering.}
We have used novel ideas for sampling inputs based on clustering on features in the input and 
features in the generated output (by some top-ranked program). The work on synthesis with abstract
examples~\cite{DBLP:conf/cav/Drachsler-Cohen17} is based on a similar intuition. It recognizes
that certain input-output examples are similar enough to be clustered and presented to the user as
one {\em{abstract example}}. The goal there is to let the user effectively give a set of concrete
examples at once to the synthesizer (by validating an abstract example). In our work, we use clustering
of inputs to perform intelligent sampling of inputs. We pick a concrete input from this sample to present
to the user.

\paragraph{Distinguishing Inputs.}
The notion of significant inputs introduced here generalizes
{\em{distinguishing inputs}}, introduced in prior work on program synthesis~\cite{bitvectors,godefroid2012automated}.
An input $i$ is distinguishing if there exist two programs $p_1$ and $p_2$ that are both consistent with
the current constraints, but produce different outputs $p_1(i)$ and $p_2(i)$ on the input $i$.
Thus, $i$ distinguishes between \emph{two} programs, and hence an additional input-output constraint for $i$
eliminates either $p_1$ or $p_2$.

In this work we generalize this idea to not just two but a \emph{set} of programs, and furthermore, emphasize
that not all such inputs are equally effective for optimizing synthesis convergence in practice.
A distinguishing input is a likely candidate for being significant.
However, a significant input must also satisfy three stronger requirements:
\begin{enumerate}\itemsep=0em
    \item A significant input is $\bot$ when the active learner is confident that it has convered.
        Hence, it is possible that distinguishing inputs exist, but the active learner nevertheless
        does not pick any of them as significant.
    \item A distinguishing input disambiguates \emph{any two} programs.
        In this work, we show that to optimize the convergence of the active learner, a significant input does not
        treat all programs equal: it prioritizes highly-ranked programs and programs that disagree with the
        current candidate.
    \item In prior work~\cite{bitvectors,godefroid2012automated}, the input space is known \emph{a priori},
        such as the space of all size-$n$ bitvectors.
        This allows closed-form formulations of the input selection problem and analysis of its convergence.
        In the practically-inspired formulation of the significant input problem (\Cref{sec:significantquestions}) the input space is
        not known in closed form.
\end{enumerate}

\paragraph{Oracle-Guided Inductive Synthesis.}
Jha and Seshia recently developed a novel formalism for example-based program synthesis called \emph{oracle\hyp{}guided
inductive synthesis} (OGIS)~\cite{ogis}.
It builds on top of \emph{counterexample\hyp{}guided inductive synthesis} (CEGIS), a common paradigm for building
interactive synthesis engines~\cite{sketch}.
In OGIS, a synthesis engine has access to an \emph{oracle}, which is parameterized with the types of queries it is able
to answer.
Typical kinds of queries include class membership, counterexamples, and distinguishing inputs.

The setting studied in this work can be likened to OGIS, with the user playing the role of an oracle providing
counterexamples.
However, it differs in two major ways.
First, we define the notion of \emph{significant} inputs that try to
minimize some convergence criteria, such as the number of iterations.
Second, we consider the ``active learner'' setting where we use information gain 
to generate significant inputs and present them proactively to the user.
Both aspects improve the usability of an interactive system.

\paragraph{Predictive Interaction.}
The notion of proactively interacting with the user during a synthesis session is known as \emph{predictive program
synthesis} or \emph{predictive interaction}.
Mayer et al.~\cite{flashprog} established that any form of interaction (such as displaying a paraphrased program
candidate or presenting distinguishing inputs) improves the correctness and subjective trust in an example-based
data transformation system.
Building on their findings, in this work we investigate how particular choices of significant inputs and techniques for
selecting them impact the convergence criteria of a synthesis interaction.
Similarly, Kandel et al.~\cite{wrangler} and Peleg et al.~\cite{peleg18} present different settings of predictive
interaction, in which the proactively sought constraints and suggestions describe the subexpressions of the desired
program, as opposed to its behavior on individual inputs.

\paragraph{Active Learning.}
The process of using significant inputs to iteratively perform program synthesis -- as described
in this paper -- is an example of active learning~\cite{al:settles}.
In active learning, data is not available a priori,
but the learner queries for data that will help it converge.
The question of significant inputs -- the
next query to make -- becomes the core problem of active learning.
In fact, since the inputs on which
the synthesized program ought to work are also available in our setting, our setting falls under what is
known as pool-based active learning~\cite{mccallumzy1998employing}.
This setting has been studied for classification, version space reduction, and other classic machine learning
domains~\cite{dasgupta2005analysis,baram2004online}, but here we study it in the real-life domain 
of a string data transformation system.

\paragraph{Software Testing.}
Significant inputs relate to synthesis in the same way as test inputs relate to verification.
The goal of both is to improve confidence in the underlying artifact after these inputs have been used to perform
synthesis or verification.
Test inputs are picked so that executions on those inputs covers, for example, all possible program paths.
Significant inputs are picked so that each one (given as an input-output constraint)
eliminates a subspace of programs, and together they eliminate
(almost) all 
unintended programs.

\paragraph{Data-driven invariant learning.} There is plenty of work in learning invariants from data~\cite{ernst:icse2000,rahulsharma:sas2013,suresh:pldi2016,madhu:oopsla2018,saswat:pldi2016}, but it is mostly in the passive setting. The significant question problem arises when synthesizing invariants using active learning. Jha and Seshia~\cite{ogis} use a synthesis framework to explore theoretical bounds on learning iterations -- what we call the optimum significant questions problem -- but they do not propose any algorithmic approach, such as the information gain approach here, to achieve the theoretical bounds.

\section{Conclusion}
The last decade of work in example-based program synthesis, and industrial applications of resulting technologies, have
shown that \textbf{(a)} program synthesis in practice proceeds as an iterative interactive session, and \textbf{(b)} the
user's cognitive load and confidence in the synthesis system largely depends on the interaction interface between the
user and the system.
\emph{Proactive} resolution of intent ambiguity is paramount to delivering high-quality user experience.
In this work, we formally study the general \emph{significant questions} problem -- questions to proactively ask the user -- and use information-theoretic notion of entropy to solve it. We instantiate the general approach to develop an active program learner that is shown to minimize the number of synthesis iterations until convergence, as well as control the number of false positive and false negative examples.

While the framework of significant questions and information gain introduced here helps optimize the \emph{convergence
criteria}, it does not directly address the criteria of the user's confidence in the program and cognitive load.
Experimentally measuring the effect of different techniques for generating user interaction on the user experience is an important area of
future work.
Further exploration of active invariant learning in program synthesis by examples is also left for future work.

\small
\bibliographystyle{abbrv}   

\appendix

\end{document}